\documentclass{appolb}
\usepackage{epsfig,amssymb}

\begin{document}
\title{Mass Corrections and Neutrino DIS
\thanks{Collaborations with 
D.~Mason, F.~Olness, M.H.~Reno and W.-K.~Tung;
presented at {\it DIS2002}, Cracow, 30 April - 4 May 2002.
Work supported by the National Science Foundation under Grant
PHY-0070443.}
}
\author{Stefan Kretzer
\address{Michigan State University, East Lansing, MI 48824}
}
\maketitle
\begin{abstract}\noindent
Including the effects of the ${\cal{O}}(\gtrsim 1\ {\rm GeV})$ masses
of the charm quark, $\tau$ lepton and target nucleon in 
DIS phenomenology is discussed with applications to 
CC neutrino DIS: 
Neutrino data for $F_2$ are revisited
within the global
analysis framework. A fully differential calculation
refines the CC charm production 
process as a gate to extract $\{ s(x), {\bar s} (x) \}$. 
New results are presented for a "heavy quark" version 
of the CTEQ6 set of PDFs and for 
($\nu_{\mu} \rightarrow \nu_{\tau}$ oscillation-signal)
$\tau$ neutrino cross sections.
\end{abstract}
  
\section{Introduction}\noindent
This contribution to the proceedings gives an overview of
some recent results for deep inelastic scattering mainly with
neutrino beams interacting through charged currents. 
Neutrino data are an important component of
PDF analyses because the weak currents
single out quark flavour combinations
different from those probed by
the electromagnetic current; e.g.~a charmed particle 
detected in the final state singles 
out the strange sea. Even more,
armed with a QCD calculation of neutrino
cross-sections, event rates 
can be related to neutrino fluxes 
in cases where the neutrino flux is unknown
or its flavour composition is expected to oscillate,
e.g.\ between $\nu_{\mu}$ and $\nu_{\tau}$. 

\section{Probing QCD using Neutrino Experiments}
\subsection{Charm Mass Effects and Neutrino Data
in Global Parton Analyses}\noindent
A satisfying conciliation of neutrino and charged lepton
structure functions at modestly
low-$x\ \sim 10^{-2}$ --
addressed before in terms of the naive parton model
as a violation of the approximate ``5/18-rule'' 
-- has been a problem for a while. 
Charm production effects and charm mass dependence in the neutrino data 
have proven important and a ``physics model independent'' 
analysis \cite{pmi} has improved 
the situation over previous comparisons which corrected
for charm in a physics model dependent way.
But comparing data sets for different hard processes
for mutual compatibility necessarily
requires the data to be 
compared to a common underlying theoretical model.
A global pQCD analysis of hard scattering data provides
the most appropriate framework for this comparison.
\begin{figure}[h]
\vspace*{-0.5cm}
\epsfig{figure=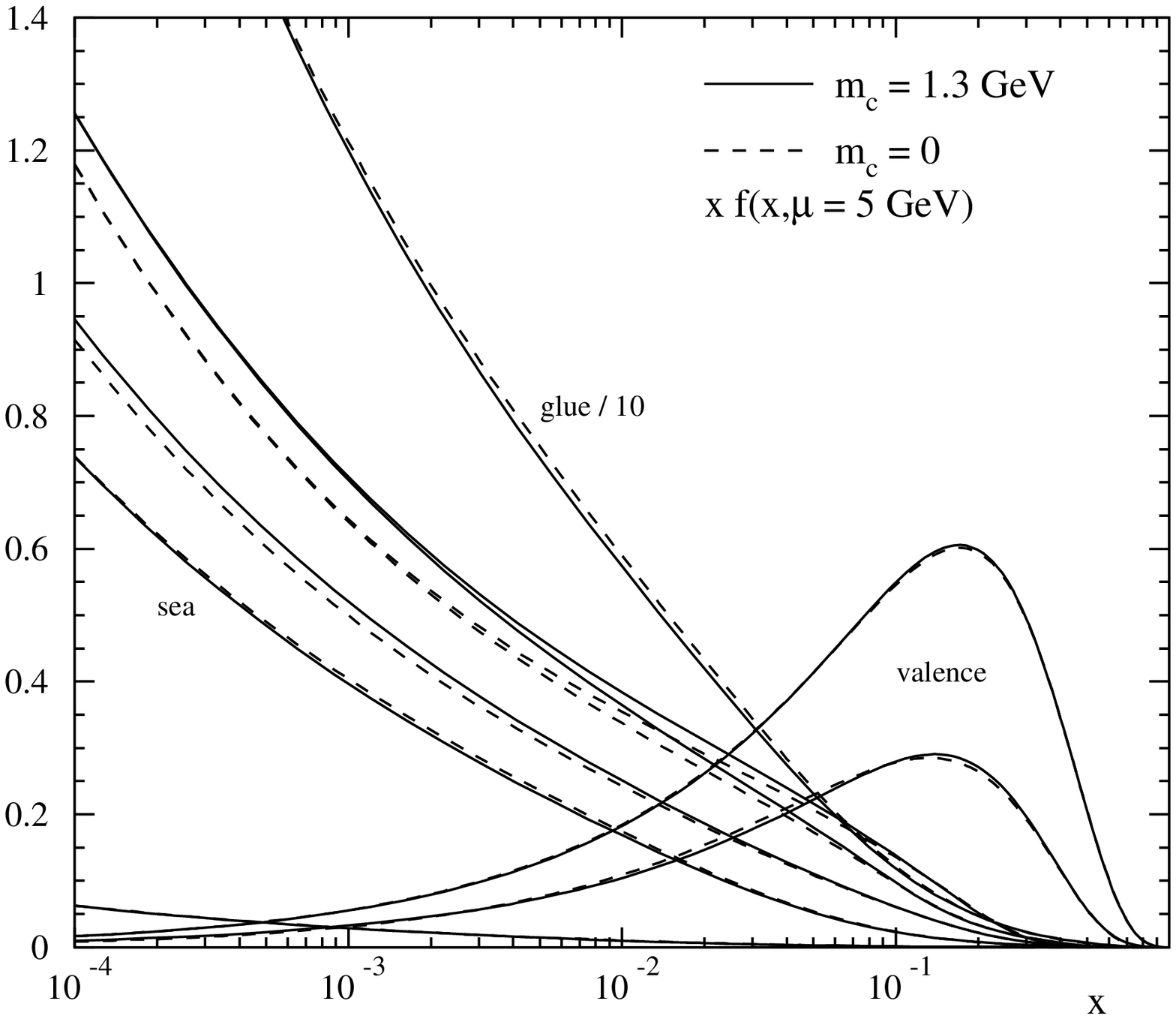,width=6cm}
\epsfig{figure=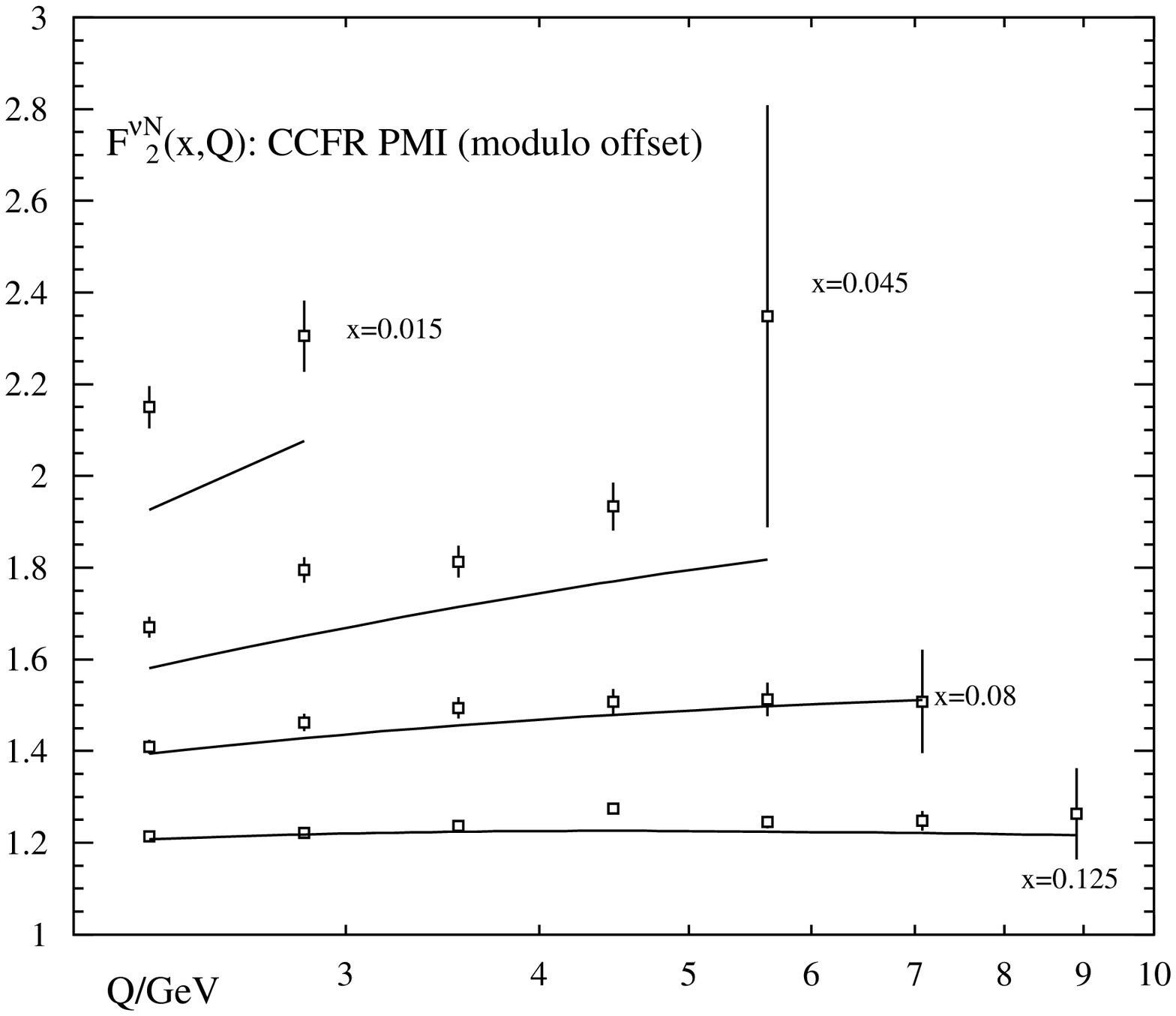,width=6cm}
\vspace*{-1cm}
\caption{
Left: The relative effect of including DIS charm threshold
effects in a 4-flavour ACOT($ \chi $) global PDF analysis.
Right: CTEQ6HQ PDF results compared with CCFR 
``physics model independent'' low-$x$ structure functions.
\label{fig:cteq}}
\end{figure}
Fig.~~\ref{fig:cteq} shows preliminary 
results of such a study \cite{kt} 
which is an extension of \cite{cteq6}. 
The PDFs in \cite{cteq6} were obtained setting $m_c=0$ in the
hard cross sections at collider energies. 
In \cite{kt} the DIS charm threshold was taken into account and
$m_c=1.3\ {\rm GeV}$ was employed with the cross sections in \cite{cprod}. 
The results in Fig.~\ref{fig:cteq} are derived from
a (simplified) ACOT ($\chi$) prescription which -- as explained in 
the second Ref.~of \cite{cprod} -- 
uses a  slow rescaling variable $\chi$
to respect
the $x$-dependent
threshold condition $W>2 m_c$; results for a PDF set 
in the {\it fixed flavour scheme} are also underway \cite{kt}. 
The left plot of
Fig.~\ref{fig:cteq} quantifies 
the amount of change in the PDFs that compensates for
introducing the charm threshold in the DIS hard scattering cross-sections.
This systematic shift can be larger than the statistical uncertainties
in the PDFs \cite{cteq6}.
In the right plot of Fig.~\ref{fig:cteq}
one observes that the agreement with the 
``physics model independent'' neutrino structure functions 
is not fully satisfactory at low-$x$.
Compare, however, with Robert Bernstein's presentation at this conference 
of preliminary NuTeV structure functions \cite{bernstein}.
The tendency seems to be that
these preliminary results compare more favourably
with the NLO PDF predictions at low-$x$.
Apart from $m_c$, there is no room here to discuss 
further theoretical factors in the evaluation of neutrino
structure functions
and the reader is referred to \cite{kosty} and to the literature
quoted therein. 
As it stands now, the data for $F_2^{\nu}$ in Fig.~\ref{fig:cteq} 
and for $\Delta xF_3^{\nu}$ as analyzed in \cite{kosty} are not
described fully satisfyingly by perturbative QCD.

\subsection{Differential Distributions for Charm in Neutrino-Production}
\noindent
The strange sea density $s(x,\mu^2 )$
is the least well determined of the
quark PDFs \cite{kosty}. Interest in $s(x,\mu^2 )$ was revived
recently also from the fact that the anomaly in the NuTeV measurement
of the Weinberg 
angle may depend on intrinsic $\left| uud s {\bar s} \right>$
fluctuations generating $(s - {\bar s})(x,\mu^2 )\neq 0$ \cite{nuano}. 
Global QCD fits have previously employed
the integrated strangeness suppression factor 
$\kappa = \int d x x (s + {\bar s} ) / \int dx x ({\bar u} + {\bar d})$
to constrain $s(x,\mu^2 )$.
More detailed information can be expected
from analyzing CC neutrino-production of charm ($W^+ s\rightarrow c$)
at fully differential level including all
NLO diagrams. As in the NC case, theory needs to provide differential 
information because of detector non-isotropy and experimental cuts.
\begin{figure}[h]
\vspace*{-2cm}
\epsfig{figure=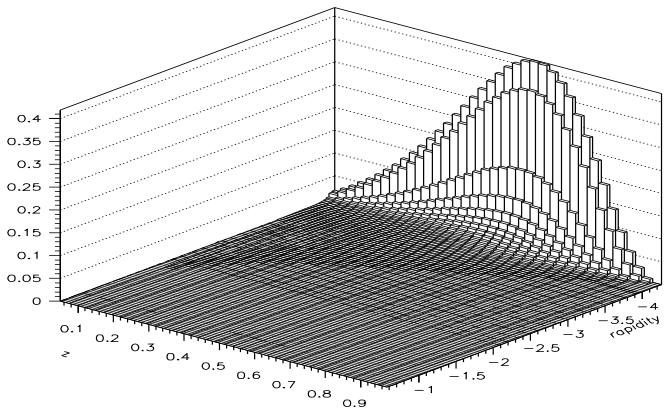,width=11cm,height=9cm}
\vspace*{-2cm}
\caption{Binned differential distribution for CC 
neutrino-production of charm on an isoscalar target: 
$E_\nu = 80\ {\rm GeV}$, $x=0.1$, $Q^2 = 10\ {\rm GeV}^2$.
\label{fig:lego}
}
\end{figure}
Fig.~\ref{fig:lego} shows a recent calculation \cite{kmo} for 
typical fixed target kinematics. A {\tt FORTRAN} 
code {\tt DISCO} has been made available and was
interfaced with the NuTeV detector Monte
Carlo. It should soon be possible
to fix the size of $s(x,\mu^2 )$ at NLO
and settle the question whether $(s - {\bar s})(x,\mu^2 )$  
is of relevant size.

\section{Probing Neutrino Oscillations using QCD
}\noindent
\begin{figure}[h]
\epsfig{figure=plot.c.eps,width=5cm,height=5cm,angle=270}
\end{figure}
\begin{figure}[h]
\vspace*{-6.5cm}
\hspace*{5cm}
\epsfig{figure=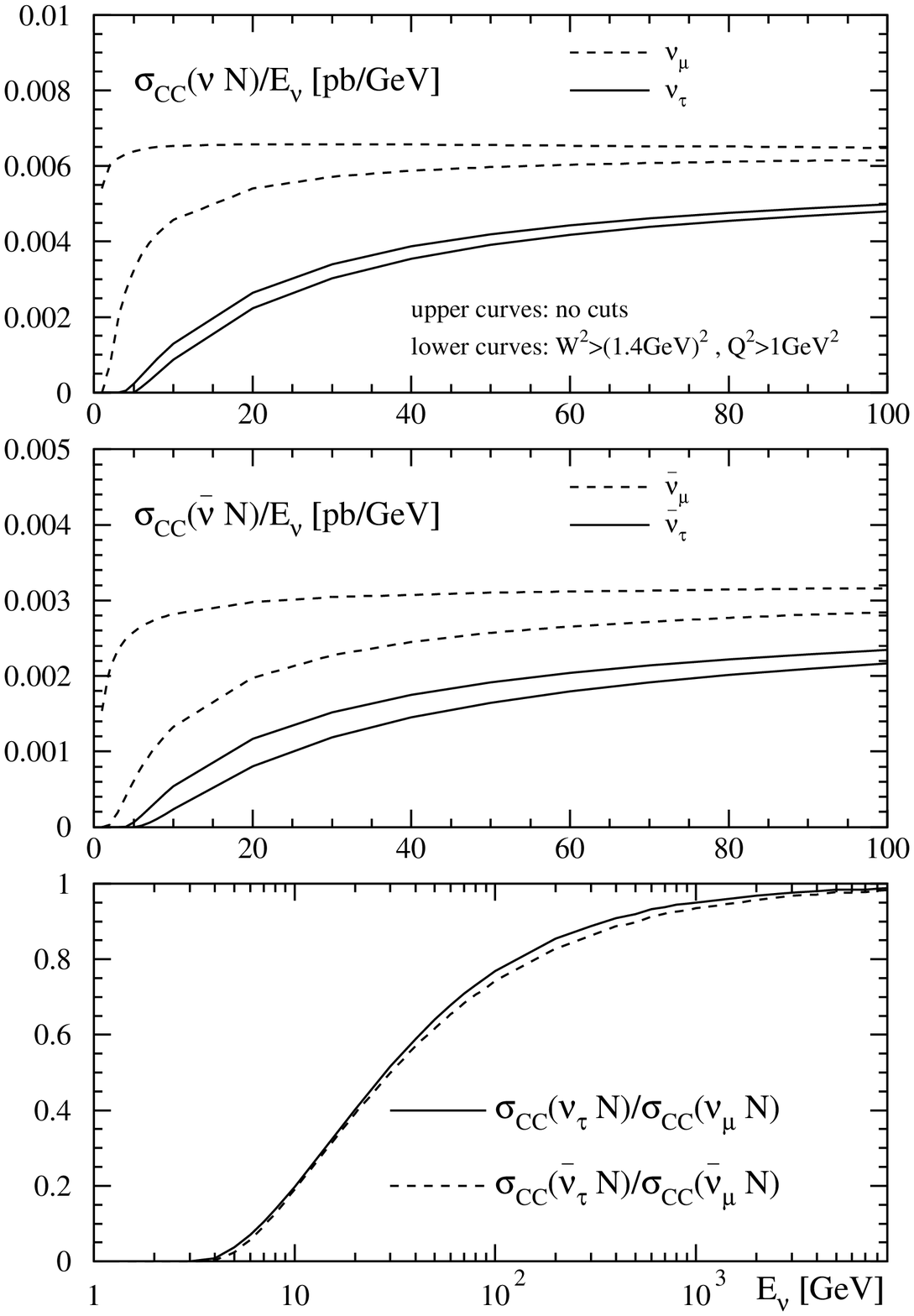,width=8cm,height=8cm}
\vspace*{-0.8cm}
\caption{Left: Violation of Eq.~\ref{eq:aj1} from mass
and NLO corrections. Right:Cross sections for inclusive neutrino- [$\sigma_{CC}(\nu N )$] 
and anti-neutrino [$\sigma_{CC}({\bar \nu} N)$]
production on an isoscalar target. 
\label{fig:tau}
}
\end{figure}
Results from the SuperKamiokande underground experiment measuring the 
atmospheric neutrino flux suggest that $\mu$ neutrinos oscillate into 
$\tau$ neutrinos with nearly maximal mixing \cite{Fukuda:1998ub}. 
A test of the oscillation
hypothesis is $\nu_\tau$ production of $\tau$
through charged current interactions, a process which will be studied
in underground neutrino telescopes
as well as long-baseline experiments
measuring neutrino fluxes from accelerator sources.
In the following the deep-inelastic contribution to $\nu_\tau N\rightarrow
\tau X$ is presented
incorporating NLO QCD corrections, target mass,
$\tau$ mass and charmed quark mass corrections \cite{kr}.
Future work will combine DIS with elastic and resonant 
neutrino-production channels.
The charged current $\nu_{\tau}$ (anti-)neutrino 
differential cross section is
represented by a standard set of 5 structure functions \cite{taudis}:
\begin{eqnarray} \nonumber
\frac{d^2\sigma^{\nu(\bar{\nu})}}{dx\ dy} &=& \frac{G_F^2 M_N
E_{\nu}}{\pi(1+Q^2/M_W^2)^2}\
\left\{
(y^2 x + \frac{m_{\tau}^2 y}{2 E_{\nu} M_N})
F_1^{W^\pm} \right. \\ \nonumber
&+& \left[ (1-\frac{m_{\tau}^2}{4 E_{\nu}^2})
-(1+\frac{M_N x}{2 E_{\nu}}) y\right]
F_2^{W^\pm}
\\ \nonumber
&\pm& 
\left[x y (1-\frac{y}{2})-\frac{m_{\tau}^2 y}{4 E_{\nu} M_N}\right]
F_3^{W^\pm} \\  \label{eq:nusig}
&+& \left.
\frac{m_{\tau}^2(m_{\tau}^2+Q^2)}{4 E_{\nu}^2 M_N^2 x} F_4^{W^\pm}
- \frac{m_{\tau}^2}{E_{\nu} M_N} F_5^{W^\pm}
\right\}\ .
\end{eqnarray} 
$F_4$ and
$F_5$ are
ignored in $\mu$ neutrino interactions because of 
a suppression
factor depending on the square of the charged lepton mass ($m_\ell$)
divided by the nucleon
mass times neutrino energy,
$m_\ell^2/(M_N E_\nu)$. At LO, in the
limit of massless quarks and target hadrons, $F_4$ and $F_5$ are
\begin{eqnarray} \label{eq:aj1}
F_4 & & =0 \\ \label{eq:aj2}
{2x F_5} & & = F_2 \ ,
\end{eqnarray}
where $x$ is the Bjorken-$x$ variable.
These generalizations of the Callan-Gross relation $F_2=2xF_1$ are
called the Albright-Jarlskog relations.
As with the Callan-Gross relations, the Albright-Jarlskog relations are
violated by NLO\footnote{Ref.~\cite{kr} finds that
Eq.~(\ref{eq:aj2}) is {\it not} violated at NLO in {\it massless} QCD.}
QCD and kinematic mass corrections.
Fig.~\ref{fig:tau} quantifies the violation of Eq.~(\ref{eq:aj1})
and compares $\nu_{\mu}$ and $\nu_{\tau}$ DIS interactions
with and without DIS cuts. The effect of these imposed cuts is much less pronounced 
for $\nu_{\tau}$ DIS where $m_{\tau}$ acts as a physical cut-off of non-DIS
interaction. It may surprise how slowly 
$\sigma_{CC}(\nu_{\tau} N)$ approaches $\sigma_{CC}(\nu_{\mu} N)$ from
below at very high neutrino energies 
indicating a persistent $\tau$ threshold effect.  
About half of the reduction at high energies
is actually of dynamic origin, to be attributed to
a negative contribution of $F_5$ to (\ref{eq:nusig}). 
Around e.g.~1 TeV the $m_{\tau}^2/E_{\nu} M_N$ 
suppression is compensated to some extent
by the low-$x$ rise of $F_5 \sim q(x)$ which is not
tempered by factors of $x$ or $y$.
The net effect is that the remaining
threshold suppression is seemingly doubled.

\end{document}